\documentclass[preprint,review,3p,12pt]{elsarticle}

\let\today\relax
\makeatletter
\def\ps@pprintTitle{%
    \let\@oddhead\@empty
    \let\@evenhead\@empty
    \def\@oddfoot{\footnotesize\itshape
         { } \hfill\today}%
    \let\@evenfoot\@oddfoot
    }
\makeatother

\biboptions{sort&compress}
\usepackage{epsfig}
\graphicspath{{Figure/}}
\usepackage{booktabs}
\usepackage{natbib}
\usepackage{amssymb}
\usepackage{graphicx,epstopdf}
\usepackage[pdf]{pstricks}
\usepackage{subcaption}
\usepackage{amsmath}
\usepackage{float}
\usepackage{xcolor}
\usepackage{tabularx}
\usepackage{verbatim}

 \usepackage{lineno}

 \makeatletter
 \providecommand{\url}[1]{%
   \begingroup
     \let\bibinfo\@Secondoftwo
     \urlstyle{rm}%
     \href{http://dx.doi.org/#1}{%
       \discretionary{}{}{}%
       \nolinkurl{#1}%
     }%
   \endgroup
 }
 \makeatother

\usepackage{hyperref}
\hypersetup{
    colorlinks=true,
    linkcolor=blue,
    filecolor=blue, 
    citecolor=blue,
    urlcolor=blue,
    pdftitle={Manuscript},
    pdfpagemode=FullScreen,
    }

\newcommand{\Dam}{Damk\(\ddot{\text{o}}\)hler number }
\newcommand{\Pec}{P\(\acute{\text{e}}\)clet number }

\begin{document}

\begin{frontmatter}



\title{Profile-Preserving Phase-Field Model for Surfactant Transport and Adsorption-Desorption in Two-Phase Flow Systems}


\author[inst1]{Haohao Hao}
\author[inst1]{Xiangwei Li}
\author[inst1]{Luyun Xu}
\author[inst1]{Tian Liu}
\author[inst1]{Huanshu Tan\corref{cor1}}
\ead{tanhs@sustech.edu.cn}
\cortext[cor1]{Corresponding author}

\affiliation[inst1]{organization={Multicomponent Fluids group, Center for Complex Flows and Soft Matter Research \& Department of Mechanics and Aerospace Engineering, Southern University of Science and Technology},
            city={Shenzhen},
            postcode={518055}, 
            state={Guangdong},
            country={China}}

\begin{abstract}
The diffuse-interface model for two-phase flows with soluble surfactants has garnered considerable attention due to its ability to circumvent the need for Robin boundary condition in the bulk surfactant transport equation.
However, the coupling between surfactant concentration and the phase field within this framework underscores the importance of accurately resolving interfacial equilibrium profiles. 
To address this limitation, we have developed a profile-preserving phase-field model for simulating surfactant transport and adsorption-desorption in two-phase flow systems.
This approach iteratively refines interfacial profiles and delta functions, removing concentration singularities and improving mass conservation.
The effectiveness of the model is demonstrated through two benchmark simulations: surfactant transport in a vortex-deformed droplet, which quantitatively reveals reduced mass error over time, and adsorption-desorption dynamics on a stationary spherical interface, showing strong agreement with one-dimensional analytical solutions for surfactant concentration distributions.
 We further highlight the model's capability by simulating the settling behavior of a surfactant-laden droplet, underscoring the critical role of adsorption-desorption kinetics in governing droplet dynamics.
\end{abstract}


%


\end{frontmatter}

\section{Introduction}
\label{sec:sample1}

Surfactant molecules, which possess both hydrophobic and hydrophilic components, readily adsorb at fluid interfaces -- especially those involving water. 
This adsorption changes the energy required to expand the interface (i.e., the interfacial surface tension), thereby exerting an inevitable influence on surface tension-driven dynamics and interfacial flows~\citep{levich1969surface,Stone_leal1990,Leal_2007,stocker2007spontaneous,Takagi2011,Manikantan2020,lohse2023surfactants}.
Because surfactants are widely used across various scientific and engineering disciplines~\citep{zhang2012effects,Perazzo2018,Banerjee2020,Tan2021,detlef2022,Michelin2023,Fernando2023,erinin2023effects,yi2024divergence}, understanding their behavior in complex fluid systems is crucial.
However, the mass transport of surfactants—including interfacial convection and diffusion, bulk convection and diffusion, as well as adsorption–desorption between the interface and the bulk—makes it challenging to experimentally measure  surfactant concentrations at both the interface and within the bulk.

To address these experimental limitations, various numerical techniques have been developed for modeling two-phase flows with surfactants.
These methods include the boundary integral method~\citep{Stone_leal1990}, the front-tracking method~\citep{ZHANG2006366,MURADOGLU20082238,BOOTY20103864,Muradoglu2014}, the immersed boundary method (IBM)~\citep{CHEN20141}, the arbitrary Lagrangian-Eulerian finite element method (ALE-FEM)~\citep{rump2024role}, the smoothed particle hydrodynamics method~(SPH)~\citep{ADAMI20101909}, the volume of fluid method~(VOF)~\citep{Pesci_Weiner_Marschall_Bothe_2018,ANTRITTER2024106231}, the level-set method~(LS)~\citep{STRICKER2017467,CLERETDELANGAVANT2017271,Xu2018}, and the phase field method~(PF)~\citep{DING20072078,SOLIGO20191292,LIU20109166}.
Additionally, hybrid approaches -- such as combining the level-set and front-tracking methods~\citep{Shin2018} or integrating the immersed boundary and immersed interface methods~\citep{HU2018201} -- have also been proposed.  
Among these, the diffuse-interface method has attracted particular attention for its simplicity and its success in modeling both insoluble and soluble surfactants, especially in two-phase flows with complex interface topologies~\citep{Teigen2009,ERIKTEIGEN2011375, JAIN2023111843,JAIN2024113277,YAMASHITA2024113292,Liu_Zhang_Ba_Wang_Wu_2020, Ba_Liu_Li_Yang_2024,farsoiya2024coupled}.


\begin{figure}[H]
  \centering
  \includegraphics[width=0.9\textwidth]{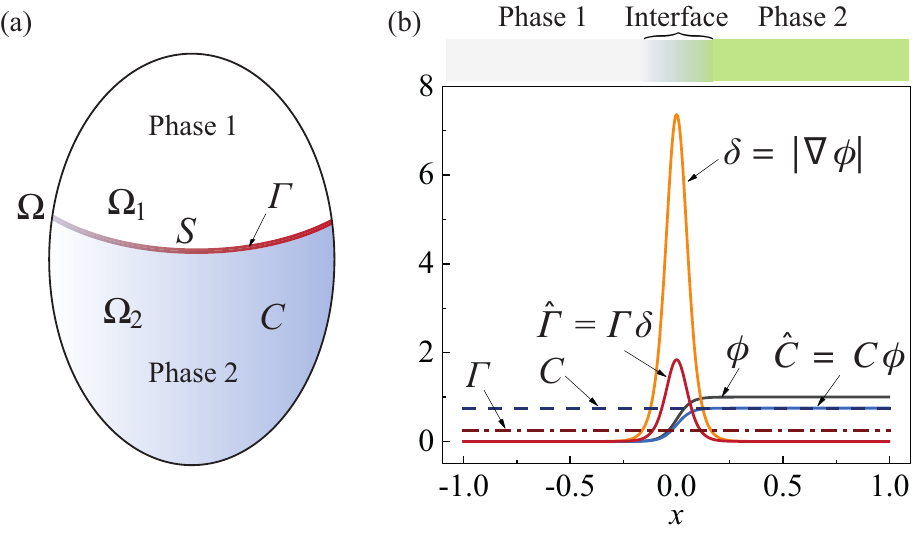}
  \caption{(a) Schematic of interfacial and bulk surfactant concentrations \(\varGamma\) and \(C\). 
  The white wine gradient color at the interface \(S\) denotes the interfacial surfactant concentration, while the white blue gradient color in \(\Omega_2\) (phase 2) represents the bulk surfactant concentration. 
  (b) Profiles of phase field function \(\phi\), surface delta function \(\delta\), interfacial and bulk surfactant concentrations \(\varGamma\) and \(C\), and their diffuse-interface forms \(\hat{\varGamma}\) and \(\hat{C}\) in one dimension. 
  For simplicity, we assume the values of \(\varGamma\) and \(C\) are constant in the normal direction of the interface.
   }
  \label{fig:mesh11}
\end{figure}

In this diffuse-interface framework, surface quantities can be directly captured on a fixed grid.
Specifically, instead of solving the interfacial surfactant transport equation~(Eq.~\eqref{STE_O1}) for the surfactant concentration $\Gamma$ on the deforming interface $S$ and the bulk surfactant transport equation~(Eq.~\eqref{bulk}) for the concentration $C$ within the evolving and irregular domain $\Omega_2$ (Fig.~\ref{fig:mesh11} (a)), the corresponding diffuse-interface formulations~(Eqs.~\eqref{STE_O2} and~\eqref{bulk_diffuse}) are solved over the entire domain $\Omega$~\citep{Teigen2009,ERIKTEIGEN2011375,Xu2018}.
This is achieved by introducing  $\hat{\varGamma}=\varGamma \delta$ and $\hat{C}=C \phi$, where $
\delta$ is a regularized delta function and $\phi$ is the phase field function (Fig.~\ref{fig:mesh11} (b)).
These models have been successfully applied to simulating interfacial flows with solute surfactants, further enhancing the method's versatility~\citep{Liu_2018,LIU20109166, SOLIGO20191292, hu2021diffuse, jainmodeling2023,JAIN2023111843,JAIN2024113277}.
Recent developments, such as the immiscible-scalars model and interface-confined model, have improved the robustness and stability of the diffuse interface method~\citep{jainmodeling2023,JAIN2023111843,JAIN2024113277}.

The diffuse-interface framework provides two notable advantages. 
First, this framework eliminates the need to enforce the Robin boundary condition~(Eq.~\eqref{Robin}) in a dynamically evolving, irregular domain when solving the bulk surfactant transport equation. 
Second, because $\hat{\varGamma}$ is defined as an Eulerian variable across the entire domain $\Omega$, it naturally handles complex topological changes of the interface.
However, a key assumption underlying this method is that the phase field function $\phi$ remains in equilibrium with the interfacial profile.
The interplay between convection and diffusion terms, along with numerical diffusion from discretization errors in the interface-capturing equation~(i.e., advective Cahn-Hilliard and Allen-Cahn equations), can cause the deviation from the equilibrium state of the interface profile in practical simulations~\citep{JACQMIN199996,HAO2024104750,soligo2019mass,chiu2019coupled}. 
The out-of-equilibrium profile of \(\phi\) can cause singularities and mass non-conservation of both interfacial and bulk surfactant concentrations, as demonstrated in Section~\ref{sec:4.2}.
These deviations can result int inaccuracies in surface tension calculations and potentially lead to unphysical numerical outcomes.

We have recently developed an enhanced profile-preserving phase-field model, integrated with a hybrid Free-energy/Continuum surface force model (FECSF), to simulate two-phase flows involving insoluble surfactants~\citep{hao2024}. 
By restoring the equilibrium interface profile at each time step~\citep{HAO2024104750}, our model ensures the assumption of interfacial profile equilibrium is consistently satisfied. 
This strategy effectively prevents surfactant concentration singularities and enhances the accuracy  of surface tension calculations and surfactant mass conservation, even when dealing with coarse mesh resolutions.

In this work, we extend our numerical model to address the transport of soluble surfactants, which is critical for a wide range of  applications~\citep{Takagi2011,zhang2012effects,Manikantan2020,Tan2021,lohse2023surfactants,yi2024divergence}.
This is achieved by integrating both interfacial and bulk surfactant transport equations within the profile-preserving Cahn-Hilliard framework.
The surfactant adsorption and desorption processes are modeled using the Langmuir isotherm. 
To calculate the surface tension forces while minimizing spurious currents, we employ the hybrid FECSF model~\citep{hao2024}.
To mitigate time-step constraints, a semi-implicit scheme is adopted for solving both interfacial and bulk surfactant transport equations, alongside the finite volume method for conservative spatial discretization. 

As a case study, we investigate the impact of surfactant solubility in the bulk phase on droplet dynamics~\citep{LAKSHMANAN_EHRHARD_2010,Muradoglu2014,Dukhin2015,Pesci2018,Castonguay2023,Jadhav_Mandal_Ghosh_2024}, with a particular focus on surfactants dissolved within droplets -- a process that has received limited attention.
Specifically, we examine the effects of the Biot and \Dam numbers on the settling behavior of a surfactant-laden droplet, along with the resulting impact on surfactant distribution and Marangoni stress at the interface.

The structure of this paper is as follows. 
Section \ref{sec:sn} briefly introduces the numerical model.
Section \ref{sec:nd} describes the numerical methods for solving the interfacial and bulk surfactant transport equations, along with the solution procedure. 
Section \ref{sec:nt} provides two test cases to evaluate the accuracy of the numerical algorithm and model performance. 
In Section \ref{sec:s5}, we present a demonstration with a particular emphasis on the influence of the Marangoni effects. 
Finally, we conclude the paper in Section \ref{sec:con}. 


\section{Methodology}\label{sec:sn}

\subsection{Modeling Surfactant Transport with Adsorption and Desorption}\label{surfactant}

The transport of soluble surfactants, which undergo an adsorption-desorption process, is modeled by considering both the interfacial concentration $\varGamma(\mathbf{x},t)$ and the bulk  concentration $C(\mathbf{x},t)$.
To frame the problem within a dimensionless context, we nondimensionalize the concentrations by the critical micelle concentration ${C}^\ast_{CMC}$ for the bulk and the corresponding saturation interfacial concentration $\varGamma^\ast_{\infty}$ for the interface, with $\ast$ denoting dimensional quantities. 

The evolution of interfacial surfactant concentration is described by a transport equation on a deforming interface~\citep{stone1990,Leal_2007}, coupled with the fluid velocity field $\mathbf{u}$.
To get its diffuse-interface formulation, we follow the approach proposed by~\cite{Teigen2009,ERIKTEIGEN2011375,Xu2018}.
Specifically, we employ $\hat{\varGamma }={\left| \nabla \phi  \right|}\varGamma$ (Fig.~\ref{fig:mesh11}b), where $\phi$ is the phase field function tracking the evolving interface~\citep{hao2024}.
Within this framework, the interfacial surfactant transport equation, incorporating  the Langmuir adsorption isotherm~\citep{kralchevsky1997chemical}, is expressed as 
\begin{equation} \label{ISTE}
  \frac{\partial  {\hat{\varGamma }}}{\partial t}+\nabla \cdot \left(\mathbf{u} \hat{\varGamma } \right)= \frac{1}{Pe_{\varGamma}}  \nabla \cdot \left[ \nabla \hat{\varGamma }-\frac{(1-2\phi )\hat{\varGamma }}{\sqrt{2}Cn }\frac{\nabla \phi }{\left| \nabla \phi  \right|} \right]+\left| \nabla \phi  \right|j,
\end{equation}
where the net flux term $j$ represents the mass transfer between the interface and the bulk, and is given by the relation,
\begin{equation}\label{MTT}
  j(\varGamma, C_s) =Bi(k{{C}_{s}}(1-\varGamma)-\varGamma),
\end{equation}
with $C_s$ representing the surfactant concentration in the sublayer of the bulk phase.
The Biot number $Bi=k_dL/U$ characterizes the ratio of the interfacial convection time scale $L/U$ to the desorption time scale $k_d^{-1}$ with the characteristic length $L$ and the characteristic velocity $U$ of the system, while the parameter $k=k_a{C}^\ast_{CMC}/k_d$ denotes the ratio of the desorption time scale $k_d^{-1}$ to the adsorption time scale $(k_a{C}^\ast_{CMC})^{-1}$. 
At the initial time ($t=0$), we assume that surfactants at the interface and in the sublayer are in equilibrium, implying a zero net flux ($j=0$). This equilibrium condition yields the relationship 
\begin{equation}\label{equi}
  \varGamma_{0} = \frac{k {{C}_{0}}}{1+k {{C}_{0}}},
\end{equation}
where the subscript ${0}$ denotes the equilibrium value.

The bulk surfactant concentration $C(\mathbf{x},t)$ evolves according to the advection-diffusion equation, which is solved in an evolving and irregular domain. 
To avoid explicitly tracking domain boundaries or implementing Robin boundary conditions (as detailed in \ref{sec:sample:appendix}), we use the diffuse-interface formulation.
Specifically, we define $\hat{C} = C\phi$ (Fig.~\ref{fig:mesh11}b), following the method in~\cite{JAIN2023111843,ERIKTEIGEN2011375}. 
Thus, the dimensionless advection–diffusion equation is
\begin{equation}\label{BSTE}
  \frac{\partial  \hat{C} }{\partial t}+\nabla \cdot \left( \mathbf{u} \hat{C}\right)=\frac{1}{P{{e}_{C}}}\nabla \cdot \left[ \nabla \hat{C}-\frac{(1-\phi )\hat{C }}{\sqrt{2}Cn }\frac{\nabla \phi }{\left| \nabla \phi  \right|}  \right]-Da\left| \nabla \phi  \right| j,
\end{equation}
where the Damk\(\ddot{\text{o}}\)hler number $Da = {\varGamma}^\ast_{\infty} / ({C}^\ast_{CMC}L)$ characterizes the ratio of the thickness of the depleted region \({\varGamma}^\ast_{\infty} / ({C}^\ast_{CMC})\) to the characteristic length \(L\).

Therefore, we have equations \eqref{ISTE} and \eqref{BSTE} describing surfactant transport both at the interface and in the bulk.
Ensuring equilibrium in the interfacial profile (Fig.~\ref{fig:mesh11}b) is essential for accurate surface tension calculations and model stability, as the diffuse-interface formulations (Eqns.~\eqref{ISTE} and \eqref{BSTE}) are inherently dependent on this equilibrium.

\subsection{Interface Dynamics and Flow Modeling with the Profile-Preserving Method}\label{sec:21CHNS}

The surfactant transport equations (Eqns.~\eqref{ISTE} and \eqref{BSTE}) are integrated with governing flow dynamics by utilizing the profile-preserving phase-field method.
The governing equations that apply the profile-preserving phase-field method~\citep{HAO2024104750,hao2024} are as follows
\begin{equation} \label{EQ_CH1}
\frac{\partial \phi }{\partial t}+ \nabla \cdot \left(\phi \mathbf{u} \right) = \frac{1}{Pe_{\phi}} \nabla^2 \eta ,
\end{equation}	
\begin{equation} \label{EQ_PP1}
       \frac{\partial \phi }{\partial \tau }=\nabla \cdot \left\{\left[\sqrt{2}Cn \left(\nabla \phi \cdot \mathbf{n(\psi)}\right)-\frac{1}{4}\left(1-{{\tanh }^{2}}\left(\frac{\psi }{2 \sqrt{2} Cn }\right)\right)\right]\mathbf{n(\psi)}\right\},
\end{equation}	
\begin{equation} \label{EQ_MOM1}
    \ \rho(\phi)\left(\frac{\partial \textbf{u}}{\partial t}+\mathbf{u}\cdot \nabla\mathbf{u}\right)=-\nabla p+\frac{1}{Re}\left\{\nabla \cdot \left[\mu(\phi)\left(\nabla\textbf{u}+\nabla\textbf{u}^T\right)\right]+\frac{{\mathbf{F}}_{\mathbf{st}}}{Ca}\right\}+ \frac{\rho(\phi)}{Fr}\mathbf{e},\
\end{equation}
\begin{equation} \label{EQ_CT1}
    \ \nabla \cdot \textbf{u} = 0,\
\end{equation}
\begin{equation} \label{EQ_ST31}   
  \frac{\mathbf{F_{st}}}{Ca}=\frac{6\sqrt{2} \eta \nabla \phi }{Ca Cn}\left[1+E \ln (1-\varGamma)\right]+ Ma{{\nabla }_{s}}\ln (1-\varGamma) \left| \nabla \phi  \right|.
\end{equation}
To capture the interfacial evolution, we apply the advective Cahn-Hilliard equation (Eqn.~\ref{EQ_CH1})~\citep{JACQMIN199996,DING20072078}.
The equilibrium interfacial profile is derived by minimizing the interfacial free energy~\citep{Cahn1959}, yielding the equilibrium form
$\phi^e(\mathbf{x},t) =\frac{1}{2}\left\{1+\tanh \left[\psi (\mathbf{x},t)/(2\sqrt{2}Cn)
  \right]\right\}$  for the order parameter $\phi(\mathbf{x},t)$, where $\psi (\mathbf{x},t)$ is the signed-distance function~\citep{JACQMIN199996,DING20072078}. 
To maintain equilibrium in the interfacial profile, the profile-preserving phase field equation (Eqn.~\ref{EQ_PP1}) is solved iteratively until $\phi(\mathbf{x},\tau)$ reaches a steady state, with $\tau$ representing the iteration time.
The flow field $\textbf{u}(\mathbf{x},t)$ in the two-phase system, which include contributions from surface tension and gravity, is governed by the dimensionless Navier–Stokes equations (Eqns.~\ref{EQ_MOM1} and \ref{EQ_CT1}).
We assume a dilute surfactant concentration, such that the variation in surfactant concentration does not affect other fluid properties, except for surface tension.
The surface tension effect, incorporating Marangoni effects, is modeled using the FECSF model (Eqn.~\ref{EQ_ST31}).

The involved dimensionless parameters include P\(\acute{\text{e}}\)clet number $Pe_{\phi}$, Cahn number $Cn$, Reynolds number $Re$, Capillary  number $Ca$, Froude number $Fr$, Elasticity number $E$, and Marangoni number $Ma$.
A comprehensive description of the model can be found in \ref{sec:21CHNS}, and the detailed derivation is available in our previous work~\citep{hao2024}.

\section{Numerical Discretization and Solution Procedure}\label{sec:nd}

\subsection{Numerical Discretization}
We employ a finite-volume method on a uniform rectangular staggered mesh in space and a semi-implicit strategy to discretize governing equations in time. Specifically, the scalar fields (\(p\), \(\mu\), \(\rho\), \(\phi\), \(\psi\), \(\hat{\varGamma}\), \(\varGamma\), \(\hat{C}\), and \(C\)) are defined at cell centers, the vector fields (the velocity \(\mathbf {u}\) and the surface tension \(\mathbf{F}_\mathbf{st}\)) are defined at cell faces, and the interface normal \(\mathbf{n}\) is defined at cell vertices.
We refer the readers to~\citep{hao2024} for more details of the numerical scheme used to solve the advective Cahn-Hilliard equation (Eq.~\eqref{EQ_CH1}), the profile-preserving equation (Eq.~\eqref{EQ_PP1}), and the Navier-Stokes equation (Eqs.~\eqref{EQ_CT1} and \eqref{EQ_MOM1}). 

 We implement a semi-implicit scheme~\citep{Ascher1995} for the temporal discretization of the interfacial and bulk surfactant transport equations. In this scheme, the diffusion terms (i.e., \(\nabla^2 \hat{\varGamma}\) and \(\nabla^2 \hat{C} \)) are discretized using a Crank-Nicolson scheme, while the other terms are discretized using a second-order Adams-Bashforth scheme. The discretized form of interfacial surfactant transport equation is formulated as follows 
\begin{equation}\label{TS_STi}
\begin{aligned} 
    &\frac{{3\hat{\varGamma }^{n+1}}-4{\hat{\varGamma}^{n}}+{\hat{\varGamma }^{n-1}}}{2\Delta t}=-\left[\frac{3}{2}\mathcal{W} (\hat{\varGamma }^{n}, \mathbf{u}^n, \phi^{n+1})-\frac{1}{2}\mathcal{W} (\hat{\varGamma }^{n-1}, \mathbf{u}^{n-1}, \phi^{n})\right]+\\
    &\qquad \qquad \qquad \qquad \qquad  \frac{1}{2Pe_{\varGamma}}\left[(\nabla^2 \hat{\varGamma})^{n+1}+(\nabla^2 \hat{\varGamma})^{n}\right],
\end{aligned}
\end{equation}
where 
\begin{equation}
  \begin{aligned} 
    &\mathcal{W} (\hat{\varGamma }^{n}, \mathbf{u}^n, \phi^{n+1})=\nabla \cdot \left( \mathbf{u} \hat{\varGamma } \right)^{n}+ \frac{1}{Pe_{\varGamma}}  \nabla \cdot \left\{\hat{\varGamma}^n \left[\frac{(1-2\phi)}{\sqrt{2}Cn }\frac{\nabla \phi }{\left| \nabla \phi  \right|}\right]^{n+1}\right\}-\\ 
    &\qquad \qquad \qquad \qquad  \left| \nabla \phi  \right|^{n+1} j^n,
\end{aligned}
\end{equation}
and \(\Delta t\) is the time step.
Similar to~Eq.~\eqref{TS_STi}, the discretized form of bulk surfactant transport equation can be expressed as
\begin{equation}\label{TS_STb}
\begin{aligned} 
&\frac{{3\hat{C }^{n+1}}-4{\hat{C}^{n}}+{\hat{C }^{n-1}}}{2\Delta t}=-\left[\frac{3}{2}\mathcal{G} (\hat{C }^{n}, \mathbf{u}^n, \phi^{n+1})-\frac{1}{2}\mathcal{G} (\hat{C }^{n-1}, \mathbf{u}^{n-1}, \phi^{n})\right]+\\
&\qquad \qquad \qquad \qquad \qquad  \frac{1}{2Pe_{C}}\left[(\nabla^2 \hat{C})^{n+1}+(\nabla^2 \hat{C})^{n}\right],
\end{aligned}
\end{equation}
where
\begin{equation} 
  \begin{aligned}
&\mathcal{G} (\hat{C }^{n}, \mathbf{u}^n, \phi^{n+1})=\nabla \cdot \left( \mathbf{u} \hat{C} \right)^{n}+ \frac{1}{Pe_{C}}  \nabla \cdot \left\{\hat{C}^n \left[\frac{(1-\phi)}{\sqrt{2}Cn }\frac{\nabla \phi }{\left| \nabla \phi  \right|}\right]^{n+1}\right\}-\\ 
&\qquad \qquad \qquad \qquad  Da\left| \nabla \phi  \right|^{n+1} j^n.
\end{aligned}
\end{equation}

To evaluate fluxes at the cell faces for the advection terms in Eqs.~\eqref{TS_STi} and~\eqref{TS_STb}, we employ a fifth-order weighted essentially non-oscillatory (WENO) scheme~\citep{LIU1994200} with the local flow velocity determining the upwind direction. For other terms in these two equations, we apply a second-order central difference scheme to approximate the gradient operator \(\nabla \), and a second-order linear interpolation scheme to approximate the values of variable at cell face.

\subsection{Solution Procedure}

The comprehensive solution procedure for each time step loop can be summarized as follows

\begin{enumerate}
\item[(1)] Advance the phase field function \(\phi(\mathbf{x},t)\) using Eqn.~\eqref{EQ_CH1} after initialization. 
\item[(2)] Correct the profile of the phase field \(\phi(\mathbf{x},t)\) using the profile-preserving equation Eqn.~\eqref{EQ_PP1} and update the phase field function \(\phi(\mathbf{x},t)\).
\item[(3)] Compute the mass transfer term \(j(C_s,\varGamma)\) using Eqn.~\eqref{MTT}.
\item[(4)] Solve interfacial surfactant transport equation~Eqn.~\eqref{TS_STi} and compute the interfacial surfactant concentration \(\varGamma(\mathbf{x},t)\).
\item[(5)] Solve bulk surfactant transport equation Eqn.~\eqref{TS_STb} and compute the bulk surfactant concentration \(C(\mathbf{x},t)\).
\item[(6)] Compute the capillary force and Marangoni force using Eqn.~\eqref{EQ_ST31}.
\item[(7)] Solve Navier-Stokes equation Eqn.~\eqref{TS_STi} and obtain the velocity~\(\mathbf{u}(\mathbf{x},t)\) and pressure \(p(\mathbf{x},t)\) by solving Eqns.~\eqref{EQ_CT1} and~\eqref{EQ_MOM1}.
\item[(8)] Update above functions.

\end{enumerate}   


\section{Numerical Tests}\label{sec:nt}

This section presents two numerical tests to evaluate the model's performance and algorithm accuracy.
The first test examines a drop with soluble surfactant in a single vortex, comparing  interfacial and bulk surfactant profiles with and without iterative correction of the interface profile to equilibrium. 
Following the study~\cite{MURADOGLU20082238,Teigen2009}, we then simulate  surfactant mass transfer between the interface and bulk on a stationary spherical drop.
\subsection{Transport of surfactants in a drop deformed in a single vortex}\label{sec:4.2}

We examine the effect of preserving the equilibrium of interface profile on surfactant distributions in a drop deformed by a single vortex.
Surfactant exchange between the interface and bulk is neglected. 
The vortex flow is prescribed as $u=-{{\sin }^{2}}(\pi x/L)\sin (2\pi y/L)\cos \left( \pi t/T \right)$ and $v=\sin (2\pi x/L){{\sin }^{2}}(\pi y/L)\cos \left( \pi t/T \right)$, 
with $L=1$ and $T=4$. 
A drop of radius $R=0.15$ is initially placed at $(0.5,0.75)$ in a unit square domain.
The dimensionless parameters are $\Delta t=4 \times 10^{-4}$, $L/\Delta x=256$, $Cn = 0.75\Delta x$, ${\varGamma}_{0}=0.5$, ${C}_{0}=0.5$, $Pe_{\phi}=1/Cn$, $Pe_{\varGamma}=100$, and $Pe_{C}=100$. 

\begin{figure}[H]
  \centering
  \includegraphics[width=1\textwidth]{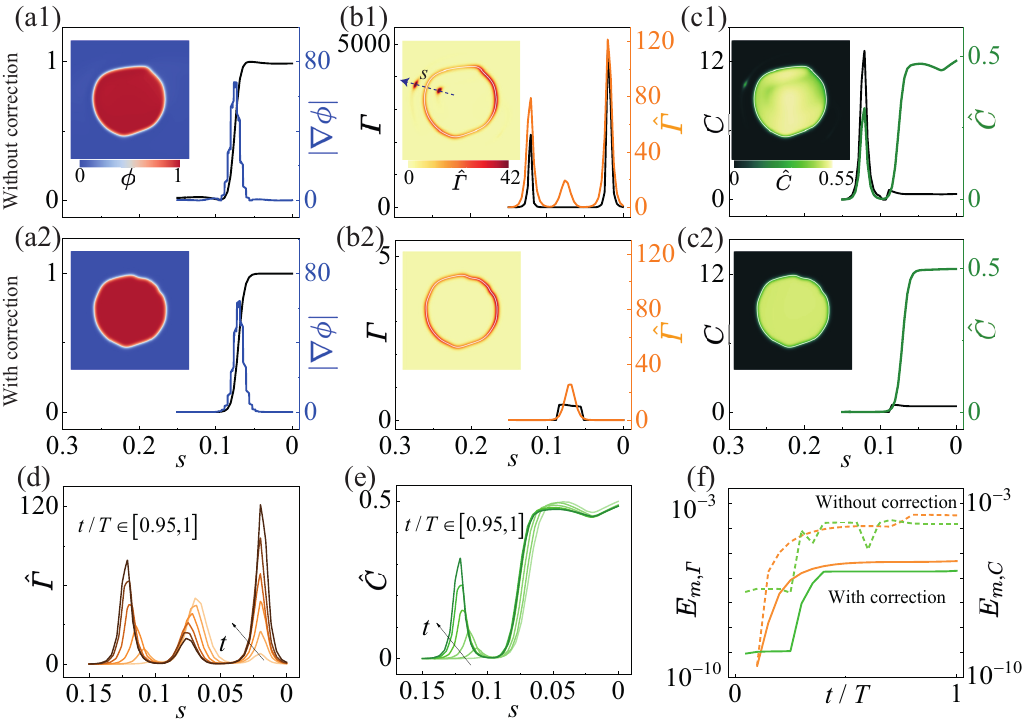}
  \caption{Surfactant transport in a vortex-deformed drop. 
 Profiles of (a) \(\phi\) and $|\nabla \phi |$, (b) $\varGamma$ and $\hat{\varGamma}$, and (c) $C$ and $\hat{C}$ after one full vortex rotation period ($t = T$).
Insets show the phase field $\phi$, interfacial concentration $\hat{\varGamma}$, and bulk concentration field $\hat{C}$. 
 For comparison, results without profile correction (a1-c1) and with correction (a2-c2) are plotted along the blue dashed line ($s$) indicated in the inset of (b1).
(d) and (e) show the temporal evolution of profiles without correction over the interval $t/T\in[0.95,1]$.
  (f) show the time evolution of mass errors of interfacial and bulk surfactant concentrations ($E_{m,\varGamma}$ in orange and $E_{m,C}$ in blue) with profile correction (solid lines) and without correction (dashed lines).}
  \label{fig:mesh4}
\end{figure}

Figure~\ref{fig:mesh4}(a1)-(c1) show the phase field \(\phi\), interfacial surfactant concentration~\(\hat{\varGamma}\), and bulk concentration~\(\hat{C}\) after one full vortex rotation period ($t = T$) without profile correction, while Figure~\ref{fig:mesh4}(a2)-(c2) show corrected results based on Eq.~\eqref{EQ_PP1}.
Profile correction maintains a drop shape closer to the initial circle by minimizing deviations of $\phi$ from equilibrium~\citep{HAO2024104750}.
Without correction, singularities appear in $\hat{\varGamma}$ and $\hat{C}$ near the interface, as highlighted along the blue dashed line ($s$) in the inset of Figure~\ref{fig:mesh4}(b1).
Applying profile correction eliminates these singularities and confines the surfactant distributions to their respective regions, restoring the expected profiles.

The origin of these singularities is analyzed via the interfacial and bulk surfactant transport equations (Eq.~\eqref{ISTE} and Eq.~\eqref{BSTE}).
Under equilibrium with zero velocity, the steady-state solutions are $\hat{\varGamma}(x) = \varGamma_0 \left\lvert \nabla \phi(x)\right\rvert $~\citep{JAIN2024113277} and $\hat{C}(x) = C_0 \phi(x)$~\citep{JAIN2023111843}, indicating that accurate surfactant profiles rely on preserving phase field equilibrium.
However, as shown in Figure~\ref{fig:mesh4}(a1), discretization errors disturb the interface, creating spurious gradients that drive artificial surfactant accumulation (Figs.~\ref{fig:mesh4}(d) and (e)) and singularity formation. 
These inaccuracies distort $\hat{\varGamma}$ and $\hat{C}$ (black lines in Figs.~\ref{fig:mesh4}(b1) and (c1)), further leading to numerical errors in surfactant mass transfer (Eq.~\eqref{MTT}) and surfactant tension forces (Eq.~\eqref{EQ_ST31}).
Profile correction effectively suppresses these errors (Figs.~\ref{fig:mesh4}(b2) and (c2)).
The temporal evolution of mass errors ($E_{m,\varGamma}$ and $E_{m,C}$) is shown in Figure~\ref{fig:mesh4}(f), where $E_{m,\varGamma}(t) = \int_{\Omega }{| {{\hat{\varGamma}(t) }}-{{\hat{\varGamma} }_{0}} |}dV / \int_{\Omega }{{\hat{\varGamma} }_{0}}dV$ and $E_{m,C}(t) = \int_{\Omega }{| {{\hat{C}(t) }}-{{\hat{C} }_{0}} |}dV / \int_{\Omega }{{\hat{C} }_{0}}dV$.
With profile correction (solid lines), mass errors (\(\sim10^{-6}\)) are two orders of magnitude smaller than without correction (\(\sim10^{-4}\)), indicating negligible surfactant mass loss and the improvement of our method~\citep{ERIKTEIGEN2011375,Xu2018,Liu_Zhang_Ba_Wang_Wu_2020}.

\subsection{Surfactant mass transfer between interface and bulk on a sphere}

\begin{figure}[H]
  \centering
  \includegraphics[width=1.0\textwidth]{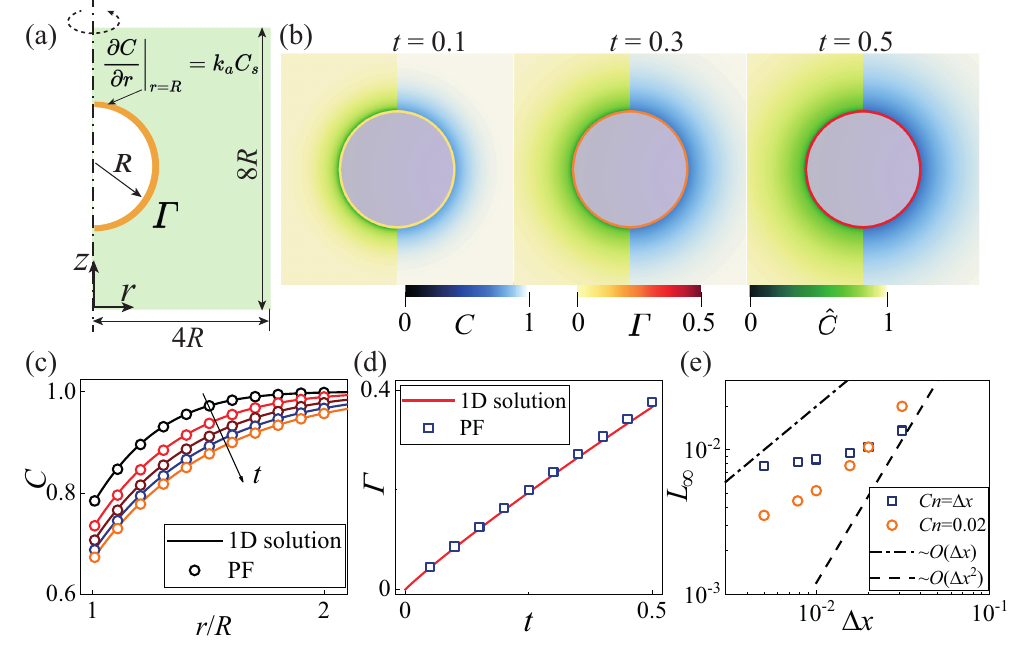}
  \caption{Surfactant mass transfer between interface and bulk on a stationary spherical drop. 
  (a) Schematic representation of surfactant adsorption on a spherical interface.
  (b) Numerical snapshots of distributions of \(\varGamma\) at the interface, and bulk surfactant concentration with \(\hat{C}\) shown on left side and \(C\) on the right side at \(t=0.1\), 0.3, and 0.5.
  (c) Comparison of profiles of \(C\) (dots)  with 1D solution (solid lines) at times \(t=0.1\), 0.2, 0.3, 0.4, and 0.5. 
  (d) Comparison of temporal evolution of \(\varGamma\) with 1D solution (solid lines). 
  (e) Convergence analysis of the infinity norm error~\(L_{\infty}\) at \(t = T\) with respect to the interface thickness~(\(Cn=\Delta x\)) and the spatial resolution~(\(Cn=0.02\)).}
  \label{fig:mesh42}
\end{figure}

To assess the accuracy of the diffusion and source terms in the surfactant transport equations (Eqs.~\eqref{ISTE} and~\eqref{BSTE}), we simulate surfactant mass transfer between the interface and bulk on a sphere of radius $R$~\cite{MURADOGLU20082238}.
The initial conditions assume a uniform bulk concentration ${{\left. C(r,t) \right|}_{t=0}}={{C}_{0}}$ and a clean interface $\left.\varGamma(r,t)\right|_{t=0 }=0$. 
Only the adsorption term (via source term~\(j\)) is considered, with the diffusion time scale in the bulk assumed much longer than the adsorption kinetics.
Neglecting convection, the bulk surfactant concentration evolves as, 
\begin{equation} \label{421}
  \ \frac{\partial C}{\partial t}=\frac{{{D}_{C}}}{{{r}^{2}}}\frac{\partial }{\partial r}\left( {{r}^{2}}\frac{\partial C}{\partial r} \right), \ 
\end{equation}
with the boundary condition ${{\left. \frac{\partial C}{\partial r} \right|}_{r=R}}={{k}_{a}}{{C}_{s}}$ and ${{\left. C \right|}_{r=\infty }}={{C}_{0}}$.
The surfactant mass conservation equation gives the interfacial mass ${{M}_{\varGamma }}(t)$ and concentration $\varGamma (t)$ as ${{M}_{\varGamma }}(t)={{M}_{C}}(0)-4\pi \int_{R}^{4R}{C(r,t){{r}^{2}}dr}$ and $\varGamma (t)={{M}_{\varGamma }}(t)/4\pi {{R}^{2}}$~\cite{MURADOGLU20082238}.
We use a finite volume method (FVM) and explicit Euler method~\cite{Teigen2009} to discretize the diffusion equation (Eq.~\eqref{421}) and compare the result with the 1D solutions derived from the phase-field framework.
Figure~\ref{fig:mesh42}(a) shows the computational domain of size $4R \times 8R$, with the fixed sphere centered at $z=4R$. 
Parameters are chosen as: mesh size $R/\Delta x=50$, time step~\(\Delta t=5 \times 10^{-4}\), Cahn number~\(Cn=0.75\Delta x\), initial bulk  concentration~\({{C}_{0}}=1\), diffusion coefficients~\({{D}_{C}} = {{D}_{\varGamma}}=1\), and adsorption rate \({{k}_{a}}=1\). 
An axisymmetric boundary condition is applied at \(r=0\), while Neumann boundary conditions on the remaining boundaries. 

As surfactant adsorbs onto the interface, the bulk concentration near the interface decreases (Fig.~\ref{fig:mesh42}(b) and (c)). 
At the same time, surfactant diffuses toward the interface, preserving spherical symmetry in the concentration contours and driving a continuous increase in $\varGamma$ (Figs.~\ref{fig:mesh42}(b) and (d)). 
Our numerical results show good agreement with 1D solutions for both bulk concentration~\(C\) profiles and interfacial concentration~\(\varGamma\), confirming the accuracy of the numerical algorithm. 
To assess mesh convergence, we introduce the infinity norm error~\citep{Teigen2009,YAMASHITA2024113292} $L_{\infty}= \text{max}\left\lvert \varGamma(t) - \varGamma_{1\text{D}}(t)\right\rvert$, where $\varGamma_{1\text{D}}(t)$  represents the 1D analytical solution.
Figure~\ref{fig:mesh42}(e) presents convergence rates of $L_{\infty}$ at $t=0.5$ for two cases: variable numerical interface thickness (\(Cn=\Delta x\)) and fixed numerical interface thickness (\(Cn=0.02\)). 
The results indicate sub-first-order convergence for variable interface thickness, while fixed thickness achieves convergence between first and second order accuracy.

\section{Influence of Surfactant Adsorption-Desorption on Droplet Settling}\label{sec:s5}

\begin{figure}[H]
  \centering
  \includegraphics[width=0.9\textwidth]{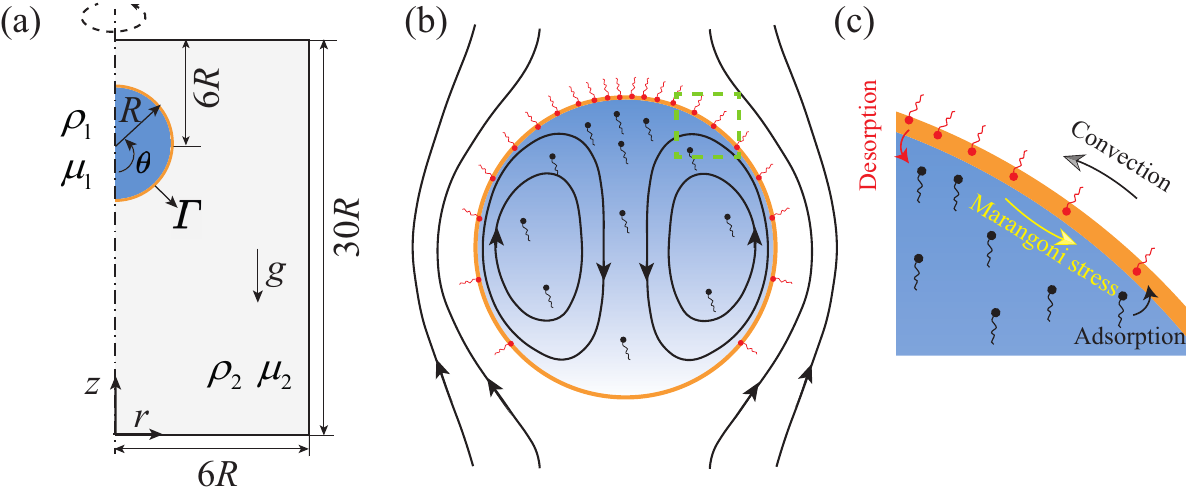}
  \caption{(a) Sketch of a surfactant-laden droplet of radius \(R\) settling due to gravity. 
  Surfactants are dissolved inside the droplet with initially uniform distribution of concentration (blue region). 
  (b) Sketch of distribution of interfacial and bulk surfactant concentrations. 
  The interfacial fluid motion convects the surfactant from the front toward the rear of the droplet, thereby establishing the gradient of interfacial surfactant concentration. 
  (c) Sketch of surfactant mass transfer through adsorption and desorption processes and Marangoni stress arsing from the interfacial surfactant gradient.
  Adsorption occurs at the front of the droplet owing to the decrease in the surfactant concentration, while desorption takes place at the rear due to surfactant accumulation.}
  \label{fig:mesh6}
\end{figure}

In this section, we apply the proposed method to investigate the settling of a surfactant-laden droplet.
The droplet, assumed axisymmetric with radius $R$, is initially placed at $z=24R$ in a cylindrical domain of radius $6R$ and height $30R$, as shown in Figure~\ref{fig:mesh6}(a).
Unlike previous studies where surfactants were dissolved in the external phase~\cite{CHEN1996144,Manikantan2020,Muradoglu2014,Castonguay2023,Jadhav2024}, we investigate a scenario where the surfactant is initially confined within the droplet but not in the external phase~(Fig.~\ref{fig:mesh6}(b)).

This setup highlights the coupling between interfacial adsorption-desorption dynamics and the convective transport inside the droplet, both influencing the droplet's settling behavior. 
As illustrated in Figure~\ref{fig:mesh6}(c), settling-induced convection transports surfactant from the front to the rear of the droplet, creating an interfacial concentration gradient, $\nabla \varGamma$, that not only generates retarding Marangoni stresses~\cite{young1959motion,Leal_2007,farsoiya2024coupled} but also drives the adsorption-desorption process.
Surfactant desorbs from the droplet surface when the interfacial concentration exceeds that in the sublayer, while adsorption occurs when the interfacial concentration is lower.
To examine how the adsorption-desorption process affects the settling motion, we vary the Biot number $Bi\in [10^{-3},10]$ and the Damk\(\ddot{\text{o}}\)hler number $Da \in [0.01, 0.5]$, assuming that adsorption and desorption timescales are identical, i.e., $(k_a{C}^\ast_{CMC})^{-1}=k_d^{-1}$, which leads to an adsorption number of $k=1$.
This assumption simplifies the analysis while preserving the core dynamics of the coupling.


We define the characteristic length as the droplet diameter, $L=2R$, and the characteristic velocity as $U=\sqrt{(1-\zeta_d)gL}$ with the density ratio $\zeta_d=\rho_2/\rho_1$, as gravity drives the settling process.
The Reynolds number is then defined via the Galileo number $Re=\sqrt{Ga}$, where $Ga = \rho_1^2 (1-\zeta_d)  gL^3/\mu_1^2$ quantifies the relative effects of gravity and viscosity and is set as $Ga=100$ in the simulation.
The hydrodynamics of the settling droplet are influenced by gravity, viscosity, and surface tension~\citep{Castonguay2023,Pak_Feng_Stone_2014,Jadhav2021,farsoiya2024coupled}.  
Other relevant hydrodynamic parameters include the Froude number $Fr=1-\zeta_d=0.2$, the Marangoni number $Ma = \mathcal{R} T \varGamma_{\infty}^\ast/(\mu_1 \sqrt{(1-\zeta_d)gL})=1$, and the Capillary number $Ca=\mu_1\sqrt{(1-\zeta_d)gL}/\sigma_0=0.1$.
The P\(\acute{\text{e}}\)clet number for interfacial and bulk surfactant transport are set to $Pe_{\varGamma}=100$ and $Pe_{C}=100$, respectively, with the viscosity ratio $\zeta_{\mu}=1$.

A far-field boundary condition at the top, a no-slip condition at the bottom, and a free-slip condition on the sides are applied (Fig.~\ref{fig:mesh6}(a)).
We assume initial uniform and equilibrium surfactant concentrations for both the interfacial and bulk phases, with $C_0=0.25$ and $\varGamma_0=0.2$ according to Eq.~\eqref{equi}. 
To ensure numerical stability and prevent non-physical negative values of surface tension, we impose a constraint on the surface tension, i.e., $\sigma (\varGamma)=\text{max} \left(\sigma_{\epsilon}, 1+E \ln \left(1-\varGamma\right) \right)$, where $\sigma_{\epsilon}=0.05$ is a threshold.
The P\(\acute{\text{e}}\)clet number for the order parameter $Pe_\phi$ is defined as $1/Cn$, where $Cn=0.75\Delta x$ is the interface thickness. 
Finally, a mesh convergence study, as demonstrated in~\ref{sec:sample:meshconvergence}, ensures the accuracy of the results.
Based on this, the mesh size is set to $2R/\Delta x=160$ for the simulations presented hereafter. 

\subsection{Effect of Biot Number}

The Biot number $Bi=(2 R /U)/k_d^{-1} $ in Equation~(\ref{ISTE}) quantifies the relative mass transfer rate of adsorption-desorption kinetics compared to interfacial advection, and the simulation results indicate that the Biot number ($Bi$) affects the settling dynamics of surfactant-laden droplets. 

\begin{figure}[H]
  \centering
  \includegraphics[width=0.9\textwidth]{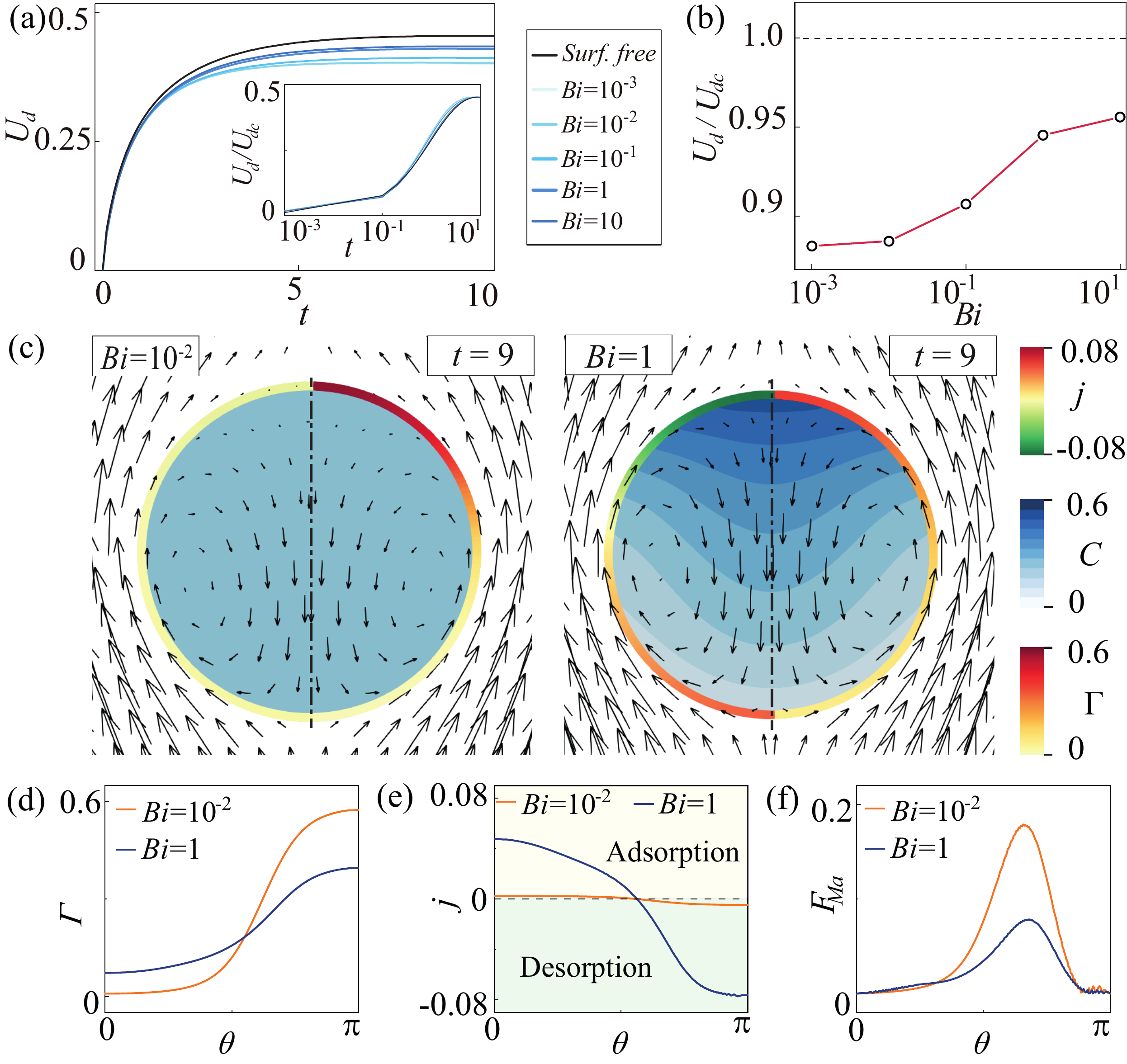}
  \caption{ (a) The droplet velocity \(U_d\) as a function of time \(t\) for different Biot numbers \(Bi\) with \(Ma = 1\), \(Da=0.1\), and \(k=1\). 
  (b) The droplet velocity ratio \(U_{ds}/U_{ds,clean}\) versus \(Bi\). \(U_{ds}\) and \(U_{ds,clean}\) denote the steady velocity of droplet with and without surfactants (clean interface), respectively.
  (c) Contours of the interfacial and bulk surfactant concentrations \(\varGamma\) (at the right half of the droplet interface) and \(C\), the adsorption-desorption flux \(j\) (at the left half of the droplet interface) , and the velocity vector (in the reference frame of the droplet) for \(Bi=10^{-2}\) and \(Bi=1\) at \(t=9\).   
  Profiles of interfacial quantities as a function of the polar angle \(\theta\) for \(Bi=10^{-2}\) and \(Bi=1\) at \(t=9\): 
  (d) Interfacial surfactant concentration \(\varGamma\), 
  (e) Adsorption-desorption flux \(j\),
  (f) Marangoni stress \(F_{Ma}\). }
  \label{fig:mesh7}
\end{figure}

Figure~\ref{fig:mesh7}~(a) shows the droplet velocity $U_d$ as a function of time $t$ for varying Biot numbers, ranging from $10^{-3}$ to $10$, while maintaining a constant Damk\(\ddot{\text{o}}\)hler number $Da = 0.1$. 
The presence of surfactants leads to a lower final settling velocity compared to the surfactant-free case (black solid line).
The evolution of settling velocity across different $Bi$ follows a similar trend, as highlighted in the inset.
However, as the Biot number increases, the droplets reach a higher final settling velocity, as shown by the variation in the normalized final velocity relative to the surfactant-free value in Figure~\ref{fig:mesh7}(b), where the dashed line represents the surfactant-free condition.


To understand how the Biot number affects the velocity reduction, we examine the relationship between $Bi$ and the changes in surfactant distribution on the droplet surface.
In Figure~\ref{fig:mesh7}(c), we present two numerical snapshots of surfactant-laden droplets at $t=9$ for $Bi=10^{-2}$ and $Bi=1$, once they have reached their final settling velocity.
The arrows indicate the flow field with the reference frame fixed on the droplet's center of mass, and the bluish shading represents the surfactant concentration, $C$.
The color coding along the droplet's left and right profiles corresponds to the adsorption-desorption flux, $j$, and the surfactant interfacial concentration, $\varGamma$, respectively.

In both cases (Fig.~\ref{fig:mesh7}(c)), the interfacial concentration is lower at the droplet's front and higher at the rear due to convection.
For $Bi=10^{-2}$, however, $\varGamma$ exhibits a steeper distribution with both a lower minimum and higher maximum concentration, as shown in Figure~\ref{fig:mesh7}(d), where $\theta$ is the angular coordinate.
This occurs because a small Biot number ($Bi=2k_d R/U \ll 1$) implies the adsorption-desorption time-scale ($k_d^{-1}$) is much larger than the advective timescale ($2R/U$), leading to a significantly weaker adsorption-desorption flux at $Bi=10^{-2}$.
Figure~\ref{fig:mesh7}(e) further reveals that adsorption occurs at the droplet's front ($\theta<\pi/2$) with positive fluxes, while desorption occurs at the rear with negative fluxes.
For small $Bi$, both adsorption and desorption rates are relatively negligible.
Consequently, the sharper gradient in $\Gamma$ induces a stronger Marangoni stress (Fig.~\ref{fig:mesh7}(f)), which retards the droplet's settling and results in a lower final settling velocity for small $Bi$, as shown in Figure~\ref{fig:mesh7}(a).
Additionally, since the disruption of the interfacial concentration has little impact on bulk concentration variation for small $Bi$, the bulk concentration becomes more uniform, as evidenced by the more homogeneous bluish color inside the droplet in Figure~\ref{fig:mesh7}(c).

For the extreme case where $Bi=0$, there is no adsorption-desorption process, which effectively corresponds to the scenario of an insoluble surfactant.
In this case, the surfactant remains at the interface, and its distribution is solely governed by the interfacial convection and diffusion without any exchange with the bulk phase~\cite{Jadhav2021,farsoiya2024coupled,FERNANDEZMARTINEZ2025105205}.

\subsection{Effect of Damk\(\ddot{\text{o}}\)hler Number}

The Damk\(\ddot{\text{o}}\)hler number $Da=({\varGamma}^\ast_{\infty} / {C}^\ast_{CMC})/(2R)$ in Equation~(\ref{BSTE}) characterizes the relative thickness of the depletion region compared to the droplet size.
A higher Damk\(\ddot{\text{o}}\)hler number indicates a thicker depletion region, which corresponds to a relatively higher surfactant capacity at the interface or a lower surfactant solubility in the bulk phase.
The simulation results confirms that the Damk\(\ddot{\text{o}}\)hler number $Da$ influences the droplet settling dynamics as well.

\begin{figure}[H]
  \centering
  \includegraphics[width=0.9\textwidth]{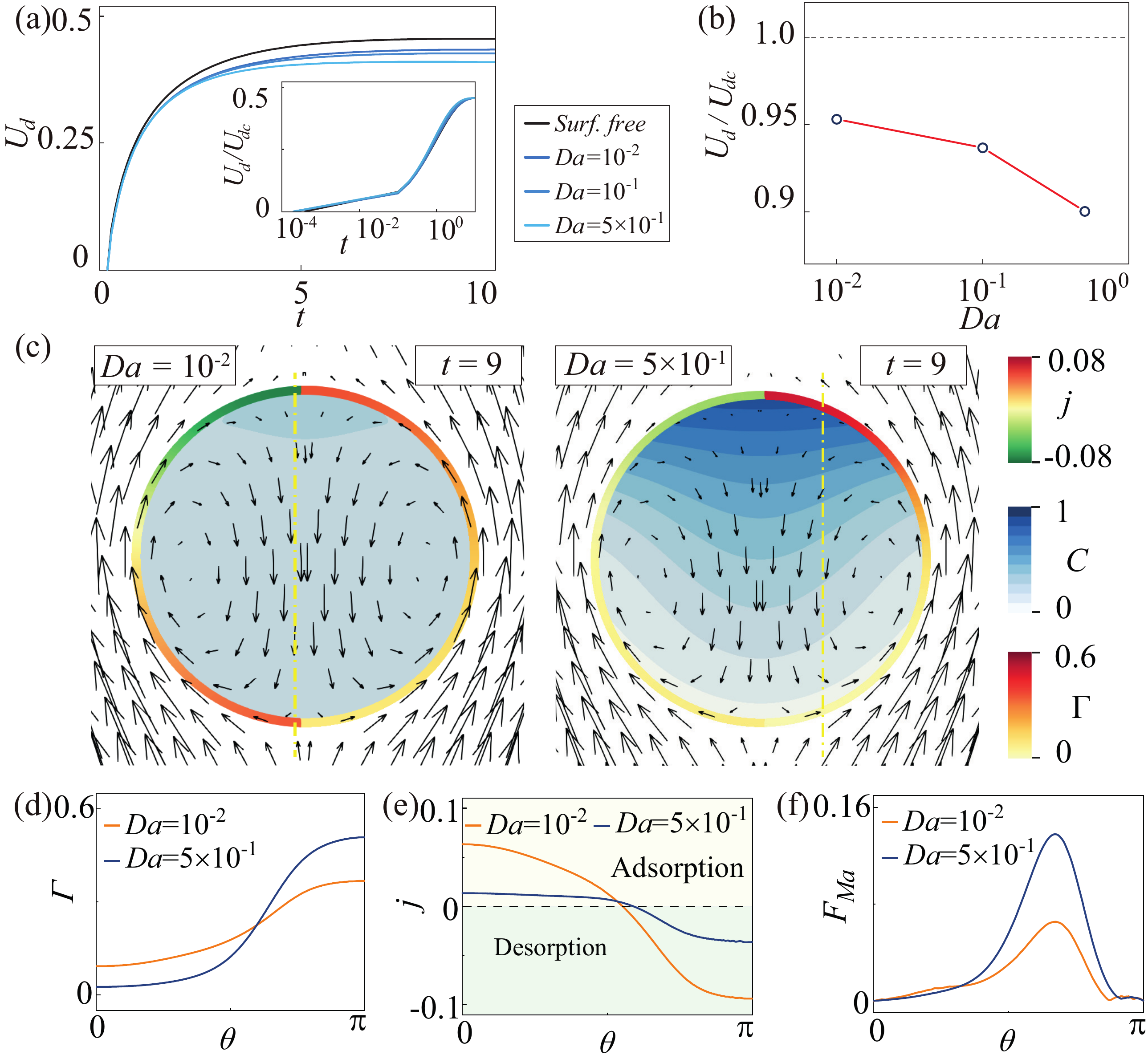}
  \caption{(a) The droplet velocity \(U_d\) as a function of time \(t\) for different Damk\(\ddot{\text{o}}\)hler numbers \(Da\) with \(Ma = 1\), \(Bi=0.5\), and \(k=1\). 
  (b) The droplet velocity ratio (\(U_d/U_{dc}\), where \(U_{dc}\) is the steady velocity of droplet without surfactant (clean interface)) versus \(Da\). 
  (c) Contours of the interfacial and bulk surfactant concentrations \(\varGamma\) and \(C\), the adsorption-desorption flux \(j\), and the velocity vector (in the reference frame of the droplet) for \(Da=10^{-2}\) and \(Da=0.5\) at \(t=9\). 
  Profiles of interfacial quantities as a function of the polar angle \(\theta\) for \(Da=10^{-2}\) and \(Da=0.5\) at \(t=9\): 
  (d) Interfacial surfactant concentration \(\varGamma\), 
  (e) Adsorption and desorption flux \(j\),   
  (f) Marangoni stress \(F_{Ma}\).}
  \label{fig:mesh9}
\end{figure}

As shown in Figure~\ref{fig:mesh9}(a), we calculated the evolution of droplet velocity by increasing the Damk\(\ddot{\text{o}}\)hler number from $0.01$ to $0.1$ and then to $0.5$, while keeping the Biot number constant at $Bi=0.5$.
For reference, the surfactant-free case (black solid line) is included in the figure. 
The normalized final velocity relative to the surfactant-free case, shown in Figure~\ref{fig:mesh9}(b), further confirms the reduction in the final settling velocity as the Damk\(\ddot{\text{o}}\)hler number increases.
To provide further insights, we present two numerical snapshots for $Da=0.01$ and $Da=0.5$ at $t=9$, when the system reaches a steady state, in Figure~\ref{fig:mesh9}(c).
The arrows, shading, and color codings are consistent with those in Figure~\ref{fig:mesh7}(c).
The simulation results reveal that a large Damk\(\ddot{\text{o}}\)hler  number ($Da=0.5$) results in a steeper distribution of the interfacial concentration $\Gamma$ (Fig.~\ref{fig:mesh7}(d)) and a corresponding weaker adsorption-desorption flux $j$ (Fig.~\ref{fig:mesh7}(e)).
This occurs because a higher $Da$ favors a greater surfactant presence at the interface, as opposed to in the bulk, implying higher interfacial capacity and lower bulk solubility.
Consequently, convection transports more interfacial surfactants toward the droplet's rear, leading to a higher interfacial concentration at the rear for $Da=0.5$.
Meanwhile, the reduced bulk solubility not only depresses the desorption flux at the rear but also limits the adsorption flux at the front, creating a larger concentration gradient and stronger Marangoni stresses (Fig.~\ref{fig:mesh7}(f)).
These effects result in a lower settling velocity in the steady state.
In contrast, for $Da=0.01$, the increased surfactant solubility in the bulk enhances both adsorption and desorption fluxes.
This causes a less steeper gradient in $\Gamma$, weakening the Marangoni stresses and ultimately leading to a slightly higher settling velocity in the steady state.

\section{Conclusion}\label{sec:con}

In summary, we have developed a diffuse-interface model for incompressible two-phase flow with sorption isotherms of dilute surfactants, utilizing the profile-preserving phase field method. 
This model maintains the profiles of the interfacial and bulk surfactants in diffuse-interface formulation since the assumption of interfacial profile equilibrium is precisely fulfilled by using the profile-preserving approach.
Numerical test involving a droplet deformed in a single vortex demonstrates the model's capability to eliminate interfacial surfactant concentration singularities while enhancing profile accuracy and mass conservation.
Further validation through simulations of surfactant mass transfer on a stationary spherical droplet verifies the model's accuracy.

We apply the model to investigate the settling of a surfactant-laden droplet.
We observe increasing the Biot number enhances adsorption/desorption fluxes, relaxing interfacial surfactant gradients. 
This reduces Marangoni stress and increases droplet velocity.
We also find increasing \Dam suppresses adsorption-desorption fluxes, amplifies interfacial surfactant gradients and Marangoni stress, and ultimately decreases steady-state droplet velocity. 

In the future, we will extend the model to incorporate surfactant micelle formation at bulk surfactant concentrations exceeding the critical micelle concentration (CMC)
In addition, at high P\(\acute{\text{e}}\)clet numbers, bulk surfactant concentration exhibits sharp gradients across small length scales \citep{BOOTY20103864}, necessitating adaptive mesh refinement (AMR) to resolve these multiscale concentration fields effectively.


\section*{Acknowledgements}
This work has received financial support from the Natural Science Foundation of China (Grants No.~12472271), and the Guangdong Basic and Applied Basic Research Foundation (Grants No. 2024A1515010509 and 2024A1515010614).

\appendix

\section{Surfactant transport equations in sharp-interface and diffuse-interface formulations}
\label{sec:sample:appendix}

The evolution of surfactant on the interface is governed by an advection-diffusion equation with a source term accounting for the surfactant mass transport between the interface and the bulk (i.e., adsorption and desorption)~\citep{Scriven1960,Leal_2007}. Its dimensionless form in sharp-interface formulation can be written as
\begin{equation} \label{STE_O1}
    \ \frac{\partial \varGamma }{\partial t}+{{\nabla }_{s}}\cdot \left( \varGamma {{\mathbf{u}}_{s}} \right)+\varGamma \left( {{\nabla }_{s}}\cdot \mathbf{n} \right)(\mathbf{u}\cdot \mathbf{n})=\frac{1}{Pe_{\varGamma}} \nabla _{s}^{2}\varGamma + j,\ 
  \end{equation}
  where \(\varGamma=\varGamma^\ast /\varGamma_{\infty}^\ast\) denotes the dimensionless surfactant concentration at the interface, \(\varGamma_{\infty}^\ast\) represents its saturation interfacial concentration, \(\nabla_s={(\mathbf{I-nn})\cdot \nabla}\) is the surface gradient operator, \(\mathbf{u_s}=(\mathbf{I-nn})\cdot \mathbf{u}\) denotes the tangential velocity along the interface, \(\mathbf{u}\) is the dimensionless fluid velocity, and \(\mathbf{n}\) is the normal direction of the interface. Here, \(Pe_{\varGamma}=UL/D_{\varGamma}\) is the \Pec representing the rate ratio between advective and diffusive surfactant transport at the interface.

  The surfactant that is soluble in one phase, is governed by the time-dependent advection-diffusion equation~\citep{Leal_2007, Manikantan2020}.  
  In sharp-interface formulation, the dimensionless form of this equation can be written as   
  \begin{equation}\label{bulk}
    \ \frac{\partial C}{\partial t}+\nabla \cdot \left( C\mathbf{u} \right)=\frac{1}{P{{e}_{C}}}\nabla \cdot \left( \nabla C \right), \
  \end{equation}
  where \(C=C^\ast /C_{CMC}^\ast\) represents the dimensionless surfactant concentration at the bulk, and \(Pe_{\varGamma}=UL/D_C\) is the P\(\acute{\text{e}}\)clet number representing the rate ratio between advective and diffusive surfactant transport in the bulk. Here, \(C_{CMC}^\ast\) is its saturation bulk concentration and \(D_C\) is the diffusion coefficient of surfactant in the bulk. The dimensionless form of the balance between surfactant diffusion to the interface and adsorption-desorption can be expressed as~\citep{Leal_2007, Manikantan2020}
  \begin{equation}\label{Robin}
    \ \frac{1}{P{{e}_{C}}}{{\left. \frac{\partial C}{\partial \mathbf{n}} \right|}_{S }}=-Da j, \
  \end{equation}
where \(Da = {\varGamma}^\ast_{\infty} / ({C}^\ast_{CMC}L)\) is the Damk\(\ddot{\text{o}}\)hler number which characterizes the ratio of the thickness of the depleted region \({\varGamma}^\ast_{\infty} / ({C}^\ast_{CMC})\) to the characteristic length \(L\).

The above surfactant transport equations (Eqs.~\eqref{STE_O1} and~\eqref{bulk}) in sharp-interface formulation can be rewritten in a diffuse-interface one, as~\citep{ERIKTEIGEN2011375}
\begin{equation} \label{STE_O2}
  \frac{\partial \left( {{\delta }}\varGamma  \right)}{\partial t}+\nabla \cdot \left( {{\delta }}\varGamma \mathbf{u} \right)=\frac{1}{Pe_{\varGamma}} \nabla \cdot \left( {{\delta }}\nabla \varGamma  \right) + \delta j,  
\end{equation}
\begin{equation}\label{bulk_diffuse}
  \frac{\partial (\phi C)}{\partial t}+\nabla \cdot \left(\phi C\mathbf{u} \right)=\frac{1}{P{{e}_{C}}}\nabla \cdot \left( \phi \nabla C \right) - Da \delta j.
\end{equation}

\section{Profile-preserving phase-field model}\label{sec:21CHNS1}

We employ the phase field method coupled with a profile-preserving approach~\citep{HAO2024104750} to implicitly capture the interface of two immiscible, incompressible fluids. This method can keep the profile of interface in a equilibrium state, maintain the mass conservation of each phase, and thus is more accurate than the tranditional phase-field method that adopts a single interface-capturing equation, e.g., the Cahn-Hilliard equation~\citep{JACQMIN199996,DING20072078}. The interface is characterized by the order parameter \(\phi(\mathbf{x},t)\), governed by the advective Cahn-Hilliard equation~\citep{JACQMIN199996,DING20072078}, as
\begin{equation} \label{EQ_CH}
    \begin{aligned}    
    &\frac{\partial \phi }{\partial t}+ \nabla \cdot \left(\phi \mathbf{u} \right) = \frac{1}{Pe_{\phi}} \nabla^2 \eta ,\\ 
    \end{aligned}
\end{equation}	
where \(\mathbf{u}\) is the dimensionless fluid velocity, \(\eta=\phi^3-1.5\phi^2+0.5\phi-Cn^2 \nabla^2 \phi\) represents the dimensionless chemical potential, and \(Pe_{\phi}=(M^*/Cn)^{-1}\) denotes the P\(\acute{\text{e}}\)clet number characterizing the ratio of convection to diffusion for the order parameter \(\phi\). 
Here, \(M^*\) is the mobility number and \(Cn=\ell/L\) is the Cahn number, where \(\ell\) is the typical interface thickness at mesoscopic scales, and \(L\) is a characteristic length scale for the macroscopic size of the fluid system. 
To attain the sharp-interface limit, where the diffuse interface converges to a sharp interface as \(Cn\) approaches zero, we employ the relations \(M^*\thicksim Cn^2\) and \(Pe_\phi \thicksim Cn^{-1} \)~\cite{magaletti2013sharp}.  
A specific form of equilibrium profiles for the order parameter \(\phi(x,t)\)~\citep{Cahn1959} can be determined by minimizing the interfacial free energy, expressed as~\citep{JACQMIN199996,DING20072078}
\(\phi(\mathbf{x},t) =\frac{1}{2}\left\{1+\tanh \left[\psi (\mathbf{x},t)/(2\sqrt{2}Cn)
  \right]\right\}\),
where \(\psi (\mathbf{x},t)\) is the signed-distance function. 


To correct the interfacial profile to maintain its equilibrium state, we use an interfacial profile-preserving approach~\citep{HAO2024104750} and the profile-preserving equation is given by
\begin{equation} \label{EQ_PP}
       \frac{\partial \phi }{\partial \tau }=\nabla \cdot \left\{\left[\sqrt{2}Cn \left(\nabla \phi \cdot \mathbf{n(\psi)}\right)-\frac{1}{4}\left(1-{{\tanh }^{2}}\left(\frac{\psi }{2 \sqrt{2} Cn }\right)\right)\right]\mathbf{n(\psi)}\right\},
\end{equation}	
where \(\tau\) is an iteration time, \(\mathbf{n(\psi)}=\nabla \psi_{\tau=0} /\left|\nabla \psi\right|_{\tau=0}\) represents the normal direction of the interface, and the signed distance function is expressed as \(\psi=\sqrt{2}Cn \ln\left[\phi/({1-\phi})\right]\).
We iteratively solve this equation with respect to \(\tau\) until the phase field function \(\phi\) satisfies the steady state criteria, defined by \( \int_{\Omega}\left|\phi_{m+1}-\phi_m\right| d \Omega \leq TOL \cdot \Delta \tau\),
where the subscript notation m-\(th\) denotes the $m$-th artificial correction step and \(TOL\) is the threshold value. 
Unless otherwise stated, we set  \(TOL=1\) and \(\Delta \tau=\text{0.005}\Delta x\), with \(\Delta x\) representing the grid size.

The motion of two-phase flows, incorporating the effects of surface tension and gravity, is governed by the dimensionless Navier–Stokes equations, expressed as follows
\begin{equation} \label{EQ_CT}
    \ \nabla \cdot \textbf{u} = 0,\
\end{equation}
\begin{equation} \label{EQ_MOM}
    \ \rho(\phi)\left(\frac{\partial \textbf{u}}{\partial t}+\mathbf{u}\cdot \nabla\mathbf{u}\right)=-\nabla p+\frac{1}{Re}\left\{\nabla \cdot \left[\mu(\phi)\left(\nabla\textbf{u}+\nabla\textbf{u}^T\right)\right]+\frac{{\mathbf{F}}_{\mathbf{st}}}{Ca}\right\}+ \frac{\rho(\phi)}{Fr}\mathbf{j},\
\end{equation}
where \(p\) represents pressure, \(\mathbf{F_{st}}\) denotes the surface tension force, and \(\mathbf{j}\) indicates the direction of gravitational acceleration. We use the Langmuir equation of state~\citep{Manikantan2020} to characterize the relationship bewteen the surface tesnion coefficient \(\sigma (\varGamma)\) and the interfacial surfantant concentration \(\varGamma\), i.e., \(\sigma (\varGamma)=1+E \ln (1-\varGamma)\), where  \(E  =\mathcal{R} T \varGamma_{\infty}^\ast/{{\sigma}_{0}}\) is the elasticity number.
Here, \(\mathcal{R}\) is the ideal gas constant, \(T\) is the temperature, and \(\sigma_0\) is the surface tension coefficient.
We adopt the hybrid Free-Energy/Continuum Surface force model~(FECSF)~\citep{hao2024}, and the surface tension force in dimensionless form is written as
\begin{equation} \label{EQ_ST3}   
  \frac{\mathbf{F_{st}}}{Ca}=\frac{6\sqrt{2} \eta \nabla \phi }{Ca Cn}\Big[1+E \ln (1-\varGamma)\Big]+ Ma{{\nabla }_{s}}\ln (1-\varGamma) \left| \nabla \phi  \right|,
\end{equation}
where \(Ma=E/Ca=\mathcal{R} T \varGamma_{\infty}^\ast/(\mu_1U)\) is the Marangoni number which quantifies the ralative strength of the Marangoni stress and the viscous stress. 

In Eq.~\eqref{EQ_MOM}, the properties of fluid 1 are characterized by density \(\rho_1\) and viscosity \(\mu_1\), while fluid 2  is defined by \(\rho_2\) and \(\mu_2\). 
Using the properties of fluid 1, we define the dimensionless fluid density \(\rho(\phi)\) and dynamic viscosity \(\mu(\phi)\) as \(\rho(\phi)=\phi +(1-\phi) \zeta_d\) and \(\mu(\phi)=\phi +(1-\phi) \zeta_{\mu}\), respectively,
where \(\zeta_d=\rho_2/\rho_{1}\) and \(\zeta_{\mu}= \mu_{2}/\mu_{1}\) represent the density ratio and the viscosity ratio, respectively. 
Three dimensionless parameters used are Reynolds number \(Re=\rho_1 U L/\mu_1\), Capillary  number \(Ca=\mu_1 U/\sigma_0\), and Froude number \(Fr=U^2/{(gL)}\), where \(g\) and \(L\) represent the gravitational acceleration and the characteristic length, respectively. 

\section{Mesh convergence tests}
\label{sec:sample:meshconvergence}

\begin{figure}[H]
  \centering
  \includegraphics[width=0.98\textwidth]{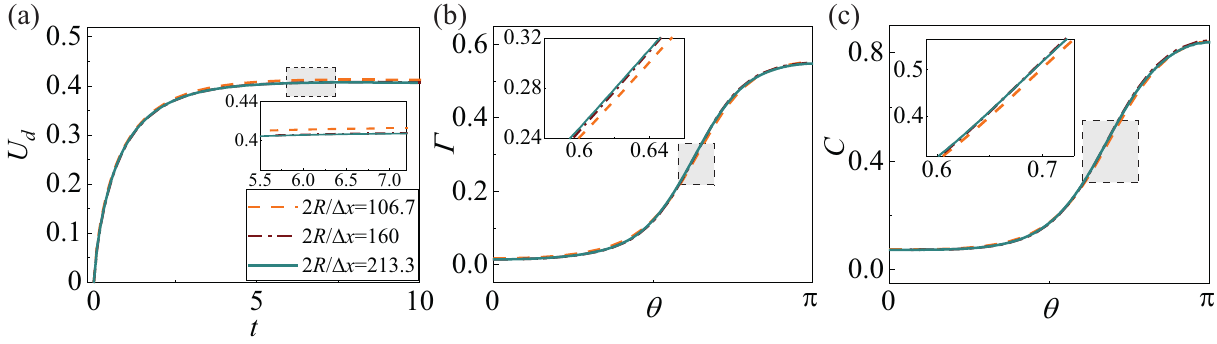}
  \caption{Mesh convergence tests for settling of a surfactant-laden droplet for \(Bi=1\) and \(Da=0.1\). (a) The droplet velocity \(U_d\) as a function of dimensionless time \(t\) with different mesh resolutions. Profiles of surfactant concentrations as a function of the polar angle \(\theta\) at \(t=9\): (b) Interfacial surfactant concentration \(\varGamma\), (c) Bulk surfactant concentration \(C\).}
  \label{fig:meshapp}
\end{figure}

We assess the accuracy of the settling velocity of the droplet and the distribution of the interfacial surfactant concentration with respect to the mesh size \(\Delta x\), as shown in Fig~\ref{fig:meshapp}.  Simulations with three different mesh sizes (\(2R/\Delta x=106.7\), 160 and 213.3) for \(Bi=1\) and \(Da=0.1\) are performed. 
From the enlarged view of Fig~\ref{fig:meshapp},
we observe the convergence behavior with increasing mesh resolution. 
We also find that the mesh size of \(2R/\Delta x=160\) ensures the accurate resolution of both interfacial and bulk surfactant concentrations.

 \bibliographystyle{elsarticle-num} 
 \bibliography{cas-refs}





\end{document}